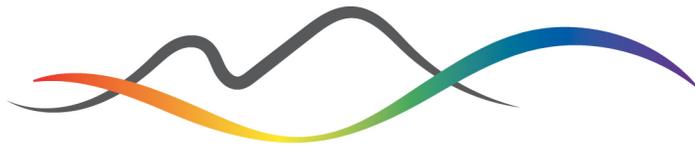

# A concise overview of the Maunakea Spectroscopic Explorer

Version 1.0, 2016-05-27

Alan W. McConnachie[1,2] (MSE Project Scientist)
*Science Team Contributors:* Carine Babusiaux[3], Michael Balogh[4], Elisabetta Caffau[3], Pat Côté[2], Simon Driver[5], Aaron Robotham[5], Else Starkenburg[6], Kim Venn[7], Matthew Walker[8]; *Project Office Contributors:* Steven E. Bauman[1], Nicolas Flagey[1], Kevin Ho[1], Sidik Isani[1], Mary Beth Laychak[1], Shan Mignot[3], Rick Murowinski[1], Derrick Salmon[1], Doug Simons[1], Kei Szeto[1], Tom Vermeulen[1], Kanoa Withington[1]

[1]*CFHT Corporation, 65-1238 Mamalahoa Hwy, Kamuela, Hawaii 96743, USA*
[2]*National Research Council of Canada, 5071 West Saanich Road, Victoria, BC, V9E 2E7, Canada*
[3]*GEPI, Obs. de Paris, PSL Research Univ., CNRS, Univ. Paris Diderot, Sorbonne Paris Cité; 5 Place Jules Janssen 92195 Meudon, France*
[4]*Department of Physics and Astronomy, University of Waterloo, Waterloo, Ontario, N2L 3G1, Canada*
[5]*ICRAR, The University of Western Australia, M468, 35 Stirling Highway, Crawley, Australia, WA 6009*
[6]*Leibniz-Institut für Astrophysik Potsdam (AIP), An der Sternwarte 16, 14482 Potsdam, Germany*
[7]*Dept. of Physics and Astronomy, University of Victoria, P.O. Box 3055, STN CSC, Victoria BC V8W 3P6, Canada*
[8]*McWilliams Center for Cosmology, Dept. of Physics, Carnegie Mellon University, 5000 Forbes Ave., Pittsburgh, PA 15213, USA*

This short document is intended as a companion and introduction to the *Detailed Science Case (DSC) for the Maunakea Spectroscopic Explorer.* It provides a concise summary of the essential characteristics of MSE from the perspective of the international astronomical community. MSE is a wide field telescope (1.5 square degree field of view) with an aperture of 11.25m. It is dedicated to multi-object spectroscopy at several different spectral resolutions in the range R ~ 2500 – 40000 over a broad wavelength range (0.36 – 1.8µm). MSE will enable transformational science in areas as diverse as exoplanetary host characterization; stellar monitoring campaigns; tomographic mapping of the interstellar and intergalactic media; the in-situ chemical tagging of the distant Galaxy; connecting galaxies to the large scale structure of the Universe; measuring the mass functions of cold dark matter sub-halos in galaxy and cluster-scale hosts; reverberation mapping of supermassive black holes in quasars. MSE is the largest ground based optical and near infrared telescope in its class, and it will occupy a unique and critical role in the emerging network of astronomical facilities active in the 2020s. MSE is an essential follow-up facility to current and next generations of multi-wavelength imaging surveys, including LSST, Gaia, Euclid, eROSITA, SKA, and WFIRST, and is an ideal feeder facility for E-ELT, TMT and GMT.

**Contacts:**
- CFHT Executive Director: Doug Simons (simons@cfht.hawaii.edu)
- MSE Project Manager: Rick Murowinski (murowinski@mse.cfht.hawaii.edu)
- MSE Project Scientist: Alan McConnachie (mcconnachie@mse.cfht.hawaii.edu)
- MSE Project Engineer: Kei Szeto (szeto@mse.cfht.hawaii.edu)

**More information:**
- The Detailed Science Case, Science Requirements Document and all original MSE science documents are available at http://mse.cfht.hawaii.edu/docs
- More information on MSE is available at http://mse.cfht.hawaii.edu

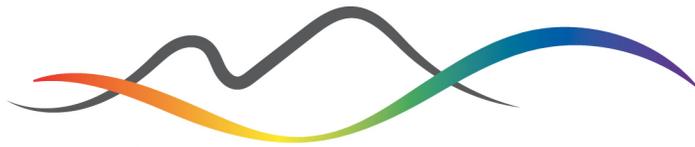

# 1 System Overview

MSE is the realization of the long-held ambition of the international astronomy community for highly multiplexed, large aperture, optical and near-infrared spectroscopy on a dedicated facility. Such a facility is the most glaringly obvious and important missing capability in the international portfolio of astronomical facilities. MSE is a refurbishment of the CFHT, upgraded to an aperture of 11.25m. Figure 1 shows the overall system architecture of MSE.

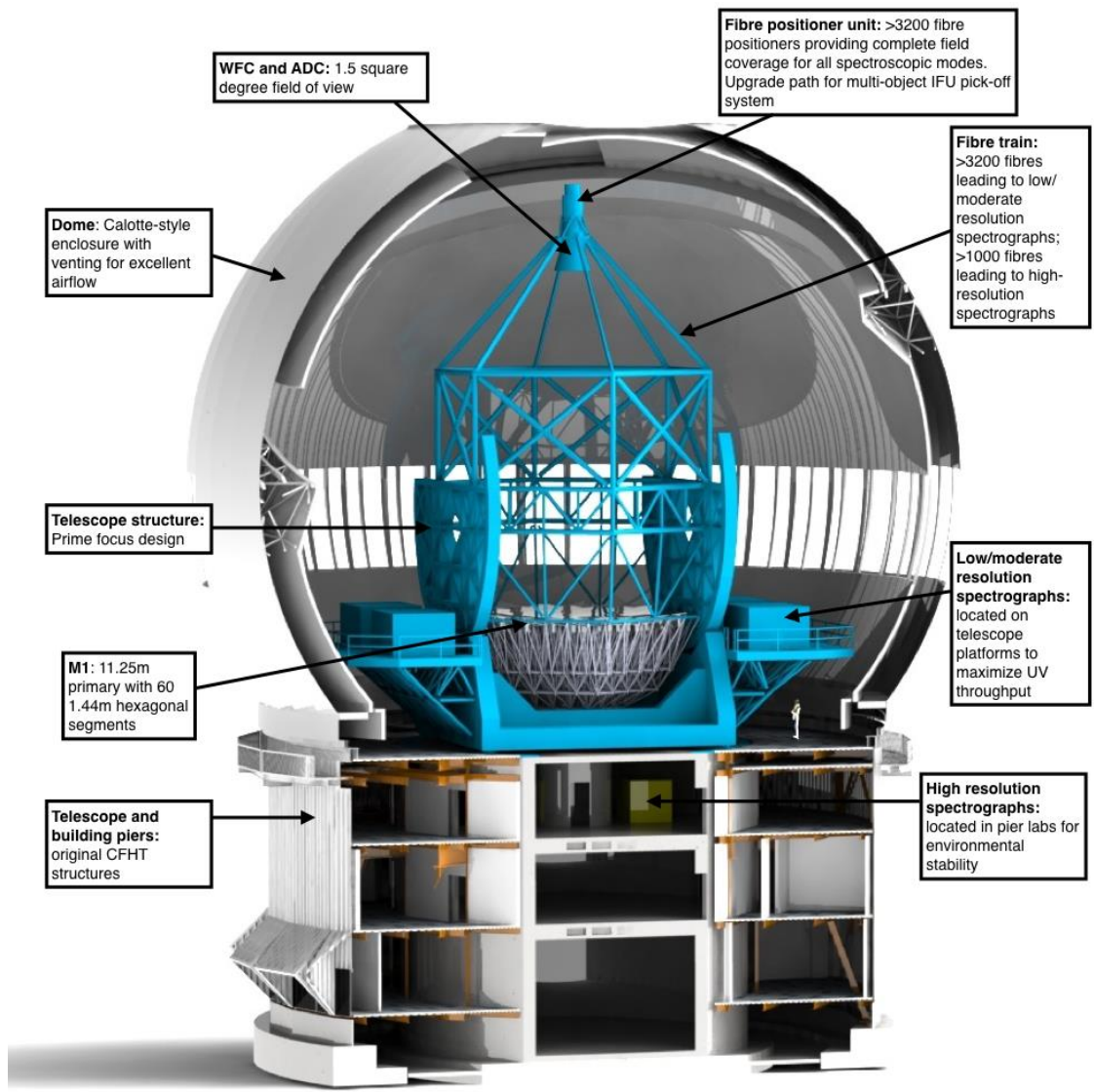

**Figure 1: Cut-away of MSE revealing the system architecture and major sub-systems.**

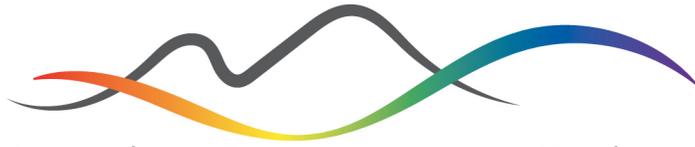

## 2 The Composition and Dynamics of the Faint Universe

### 2.1 Defining capabilities

Table 1: Summary of major science capabilities of MSE.

| Accessible sky | 30000 square degrees (airmass<1.55) | | | | | |
|---|---|---|---|---|---|---|
| Aperture (M1 in m) | 11.25m | | | | | |
| Field of view (square degrees) | 1.5 | | | | | |
| Etendue = FoV x π (M1 / 2)$^2$ | 149 | | | | | |
| Modes | Low | | Moderate | High | | IFU |
| Wavelength range | 0.36 - 1.8 µm | | 0.36 - 0.95 µm | 0.36 - 0.95 µm # | | IFU capable; anticipated second generation capability |
|  | 0.36 - 0.95 µm | J, H bands |  | 0.36 - 0.45 µm | 0.45 - 0.60 µm | 0.60 - 0.95 µm |  |
| Spectral resolutions | 2500 *(3000)* | 3000 *(5000)* | 6000 | 40000 | 40000 | 20000 |  |
| Multiplexing | >3200 | | >3200 | >1000 | | |  |
| Spectral windows | Full | | ≈Half | $\lambda_c$/30 | $\lambda_c$/30 | $\lambda_c$/15 |  |
| Sensitivity | m=24 * | | m=23.5 * | m=20.0 ♮ | | |  |
| Velocity precision | 20 km/s ♪ | | 9 km/s ♪ | < 100 m/s ★ | | |  |
| Spectrophotometric accuracy | < 3 % relative | | < 3 % relative | N/A | | |  |

♯ Dichroic positions are approximate
* SNR/resolution element = 2     ♪ SNR/resolution element = 5
♮ SNR/resolution element = 10    ★ SNR/resolution element = 30

MSE builds on the success of the SDSS concept, but is realized on a facility with ~20 times larger aperture located at possibly the world's best telescope site. Defining science capabilities include:

- **Survey speed and sensitivity:** The etendue of MSE is > 2 larger than its closest 8m competitor (149 vs. 66 m$^2$deg$^2$ for Subaru/PFS). MSE's sensitivity allows efficient observation of sub-L* galaxies to high redshift, and high resolution analysis of stars in the distant Galaxy. The excellent site image quality is essential to observe efficiently the faintest objects and to ensure the spectrograph optics are a reasonable size given the multiplexing demands.
- **Dedicated and specialized operations:** MSE's specialized capabilities enable a vast range of new science. Specialized observing modes allow time domain programs such as transient targeting, quasar reverberation mapping and precision stellar radial velocity monitoring. These require well-calibrated and stable systems only possible with dedicated facilities.
- **Spectral performance:** The extensive wavelength coverage of MSE from the UV to H-band uniquely enables the same tracers to be used to study galaxy and black hole growth at all redshifts to beyond cosmic noon. Chemical tagging with MSE can be conducted across the full luminosity range of Gaia targets, and operation at R=40000 enables the identification of weak lines in the blue to access species sampling a diverse range of nucleosynthetic sites.

### 2.2 Key science programs

The Science Requirements for MSE are defined as the capabilities necessary to conduct a suite of Science Reference Observations, high impact science programs that are *uniquely possible with MSE*. These span exoplanetary host characterization to next generation cosmological surveys. We summarize a small subset of these programs to give a flavor of the diverse science that MSE enables, and refer the reader to the DSC for a more comprehensive description.

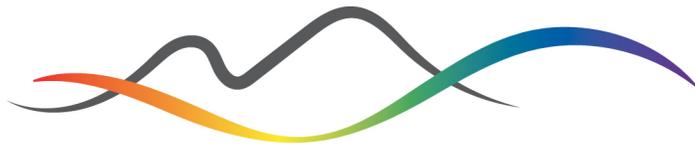

### 2.2.1 Exoplanets and stellar astrophysics

MSE will provide spectroscopic characterization at high resolution and high SNR of the faint end of the PLATO target distribution (g~16), to allow for statistical analysis of the properties of planet-hosting stars as a function of stellar and chemical parameters. With high velocity accuracy and stability, MSE time domain spectroscopic programs will allow for highly complete, statistical studies of the prevalence of stellar multiplicity into the regime of hot Jupiters for this and other samples and also directly measure binary fractions away from the Solar Neighbourhood. A menagerie of rare stellar types will be identified in MSE multi-epoch spectroscopic datasets that will allow for population studies of in-situ stars across all components of the Galaxy, at all galactocentric radii and at all metallicities.

### 2.2.2 Chemical tagging in the outer Galaxy: the definitive Gaia follow-up

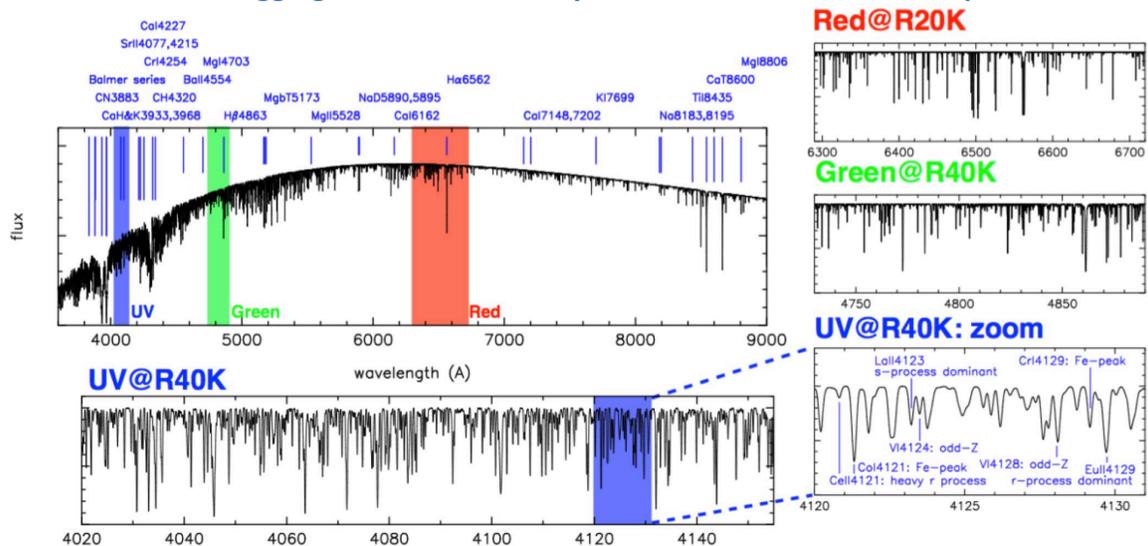

**Figure 2: Main panel shows the relative flux of a synthetic spectrum of a metal poor red giant star at the intermediate MSE spectral resolution of R~6000, along with some of the strong line stellar diagnostics accessible at this resolution. Highlighted regions show the normalized flux in three windows observable with the high resolution mode of MSE. A magnified region of the UV window shows examples of the species that will be identified at high resolution. MSE chemical tagging surveys will identify species sampling a large and diverse set of nucleosynthetic pathways and processes.**

MSE will have an unmatched capability for chemical tagging experiments. Recent work in this field has started to reveal the dimensionality of chemical space and has shown the potential for chemistry to be used in addition to, or instead of, phase space, to reveal the stellar associations that represent the remnants of the building blocks of the Galaxy. MSE will push these techniques forward to help realize the "New Galaxy", as originally envisioned by Freeman & Bland-Hawthorn (2002, ARA&A, 40, 487). MSE will focus on understanding the outer components of the Galaxy – the halo, thick disk and outer disk – where dynamical times are long and whose chemistry is inaccessible from 4-m class facilities. These components will be decomposed into their constituent star formation events by measuring abundances of chemical species that trace a large number of nucleosynthetic pathways. This includes rare species and heavy elements at blue wavelengths through R40K capabilities (Figure 2). *MSE is the only facility capable of high resolution studies of stars across the full luminosity range of Gaia targets.*

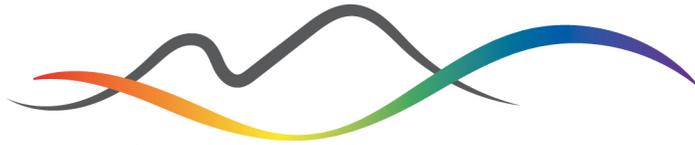

### 2.2.3  The Dark Matter Observatory

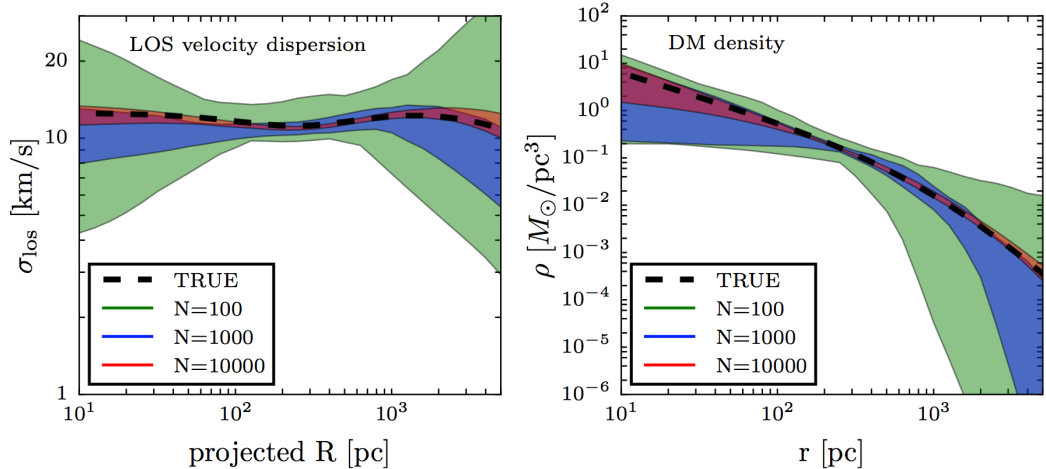

**Figure 3: Recovery of line-of-sight velocity dispersion (left) and dark matter density (right) profiles as a function of stellar spectroscopic sample size for dwarf galaxies, demonstrating the need for extensive datasets with MSE. Shaded regions represent 95% credible intervals from a standard analysis of mock data sets consisting of radial velocities for N = $10^2$, $10^3$ and $10^4$ stars (median velocity errors of 2 kms$^{-1}$), generated from an equilibrium dynamical model for which true profiles are known (thick black lines).**

MSE is the ultimate facility for probing the dynamics of dark matter over all spatial scales. For Milky Way dwarfs, MSE will obtain complete samples of tens of thousands of member stars to very large radius and with multiple epochs to remove binary stars. Such analyses will allow the internal dark matter profile to be derived with high accuracy and will probe the outskirts of the dark matter halos accounting for external tidal perturbations as the dwarfs orbit the Galaxy (Figure 3). A new regime of precision will be reached through the combination of wide field MSE *radial velocity* surveys with precision *astrometric* studies of the central regions using E-ELT, TMT and GMT. In the Galactic halo, high precision radial velocity mapping of every known stellar stream will reveal the extent of heating through interactions with dark sub-halos and place strong limits on the mass function of dark sub-halos around an L* galaxy. On cluster scales, MSE will use galaxies, planetary nebulae and globular clusters as dynamical tracers to provide a fully consistent portrait of dark matter halos across the mass function.

### 2.2.4  The connection between galaxies and the large scale structure of the Universe

Within the ΛCDM paradigm, it is fundamental to understand how galaxies evolve and grow relative to the dark matter structure in which they are embedded. This requires mapping the distribution of stellar populations and supermassive black holes to the dark matter haloes and filamentary structures that dominate the mass density of the Universe, and to do so over all mass and spatial scales. MSE will provide a breakthrough in extragalactic astronomy by linking the formation and evolution of galaxies to the surrounding large-scale structure, across the full range of relevant spatial scales (from kiloparsecs to megaparsecs).

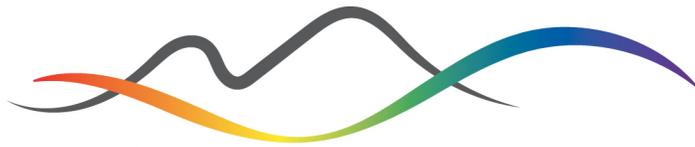

A local galaxy survey with MSE out to 100Mpc could sample our neighbourhood down to the lowest detectable masses of $3 \times 10^5$ $M_\odot$ allowing for a complete census of mass in the local Universe. A deep near infrared selected spectroscopic survey would be able to measure velocity dispersion masses for systems analogous to the local group out to z = 1, i.e., the MW, M31 and M33 would all be detectable. This will provide a direct measurement of dark matter assembly for > $10^{12} M_\odot$ halos over half the age of the Universe. The most massive halos could potentially be traced to z = 5, where the detection of any such halos would present major problems for the Cold Dark Matter paradigm. MSE will follow galaxy evolution across the peak in star-formation and merger activity, and trace the transition from merger-dominated spheroid formation to the growth of disks (Figure 4).

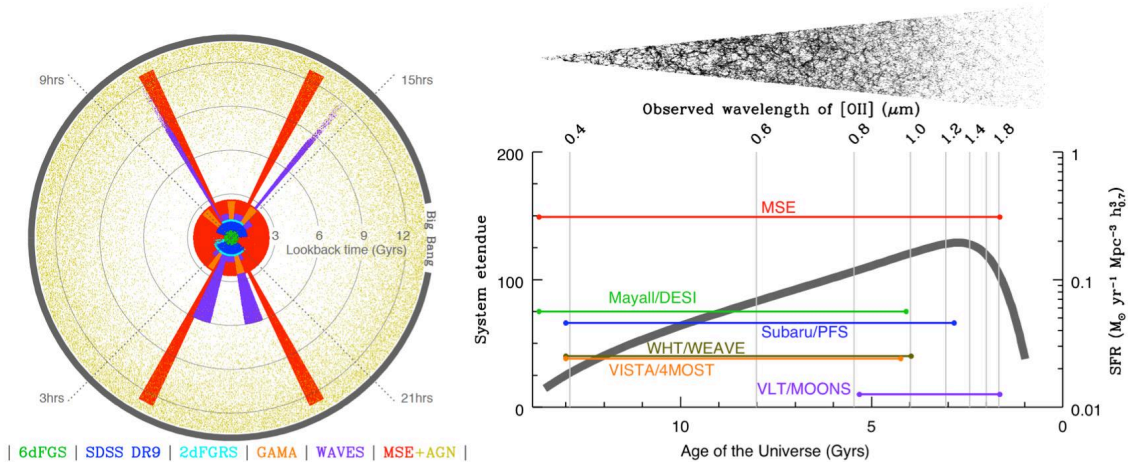

**Figure 4:** Left panel: Cone plot showing an illustrative MSE survey compared to other notable benchmark galaxy surveys. Cones are truncated at the redshift at which L* galaxies are no longer visible. Right panels: Lookback time versus cosmic star formation rate (right axis) using the parameterization of Hopkins & Beacom (2006, ApJ, 651, 142). Grey lines indicate the wavelength of OII 3727Å. Also indicated on this wavelength scale is the wavelength coverage of the major highly multiplexed spectrographs in development. These are offset according to their system etendue (left axis). The light cone demonstrates the reach of MSE for extragalactic surveys using a homogeneous set of tracers at all redshifts.

### 2.2.5 Transient targeting with MSE

Spectroscopic follow-up is essential to understanding the time-variable events discovered by LSST, SKA and other all-sky transient surveys. MSE transient studies take advantage of the dedicated operation of the facility to accumulate large datasets over time with only a small fraction of science fibres ever dedicated to targets. If N fibres in every MSE pointing observe recently identified transients (e.g., within the past 24 hours), the observing efficiency of the main MSE survey will decrease by ∼[N/3200]. However, the gain to transient science will be the equivalent of N large aperture telescopes doing nothing but transient spectroscopy with their single object spectrographs. LSST will image ∼1500 $deg^2$ per night accessible to MSE, in which there will be ~300 new supernovae and >75000 variable stars (Ridgway et al. 2014, ApJ, 796, 53). MSE will dominate the regime of faint transients and it will be able to dedicate more time to transient spectroscopy than any other large aperture telescope.

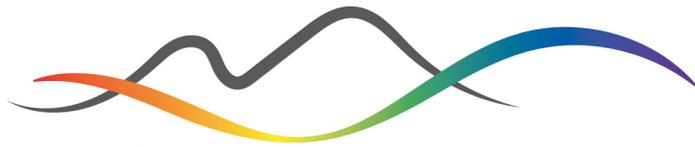

### 2.2.6 The growth of supermassive black holes

MSE will allow an unprecedented, extragalactic time domain program to measure directly the accretion rates and masses of a large sample of supermassive black holes through reverberation mapping. This information is essential for understanding accretion physics and tracing black hole growth over cosmic time. Reverberation mapping is the only distance-independent method of measuring black hole masses applicable at cosmological distances, and only ~60 local, relatively low-luminosity AGN currently have measurements of their black hole masses based on this technique. MSE will break new ground with a campaign of ~100 observations of ~5000 quasars over a period of several years (totaling ~600 hours on-sky). These observations will map the inner parsec of the quasars from the innermost broad-line region to the dust-sublimation radius. MSE will reveal the structure and kinematics in the immediate surrounding of a large sample of supermassive black holes actively accreting during the peak quasar era.

## 3 MSE and the International Network of Astronomical Facilities

Astronomy has entered the multi-wavelength realm of big facilities. Figure 5 shows some of the new observatories that collectively represent billions of dollars of investment. Each facility can operate on a stand-alone basis. However, recent history demonstrates that it is the combination of data from these facilities that will produce many of the major scientific advances. MSE occupies a critical hub in the emerging network of international astronomical facilities, which will define the future of frontline astrophysics in the 2020s and beyond.

MSE takes advantage of its equatorial location ($\ell = 19.8°N$) to access the entire northern hemisphere and more than half of the southern hemisphere (airmass < 1.55). It will share the same sky as Subaru/HyperSuprimeCam; will leverage the extensive multi-wavelength mapping of the northern skies already in-hand; will access more than half (> 10000 square degrees) of the main LSST footprint (with the ecliptic survey and any northern extensions increasing this overlap); will share two thirds of the SKA sky; will overlap significantly in accessible sky with E-ELT, TMT and GMT; will provide a ground-based complement to typically three quarters of the survey area of future space missions, including Gaia, Euclid, WFIRST, eROSITA, SPICA and PLATO.

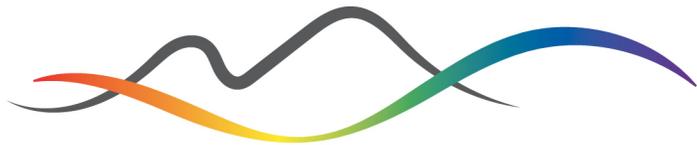

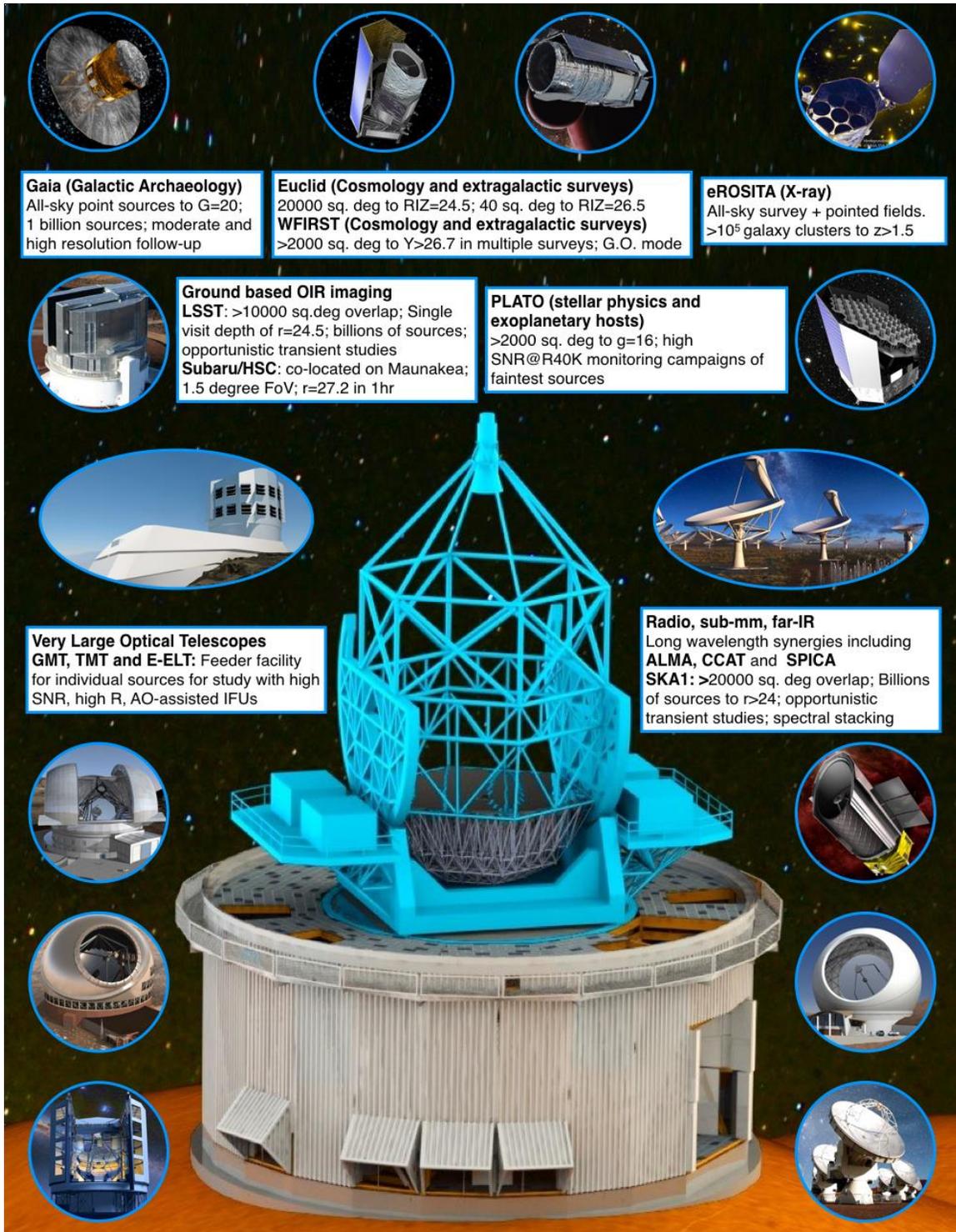

**Figure 5: MSE and other prominent astronomical facilities in the 2020s and beyond. The number and type of astrophysical targets accessible to MSE are described in the inset panels.**

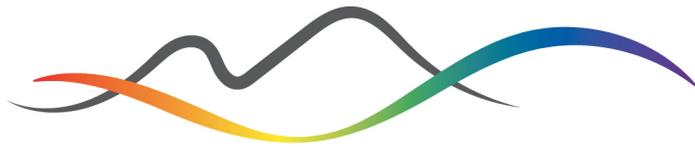

## 4 Science Operations and Data Products

MSE is a dedicated spectroscopic survey facility for operation by its partner communities. The prime deliverable is data, obtained through one or more survey programs. At first light, baseline science operations anticipate a large *legacy survey* designed by the science team (comprised of a dark – extragalactic – and bright – Galactic – component) that will occupy all available observing time. Once in standard operation, observing time will be made available for additional *strategic surveys* of flexible scope. The impact, productivity, and allocations of legacy and strategic surveys will be reviewed continually to respond to community needs and science developments.

Data will be available immediately to the survey teams that work with the Observatory to provide data products for the MSE community on a short timescale. A significant proprietary period on science data enables the MSE community to use the data for frontline science before their worldwide release. The MSE community is served by their ability to propose for, design and lead ambitious survey programs, and by a rich archive of data accessible to MSE astronomers for scientific exploitation. It is expected that > 5 million astronomical spectra will be ingested into the archive on a yearly basis (roughly equivalent to a SDSS Legacy Survey every 3 – 4 months). The scientific advantages provided to these communities are therefore significant.

The design of MSE foresees the need to upgrade components or instruments over the long lifetime of the Observatory and to respond to a changing scientific landscape. A multi-object integral field unit is already anticipated, and the top-end is designed to be upgradable to deliver this capability. MSE will remain the world's premier resource for astronomical spectroscopy: *a specialized technical capability, and a general purpose science facility*.

## 5 Development of MSE

MSE is a rebirth of the 3.6m Canada-France-Hawaii Telescope as a dedicated, 11.25m, spectroscopic facility within an expanded partnership extending beyond the original membership. The *Mauna Kea Comprehensive Management Plan* anticipates CFHT redevelopment. MSE will reuse the current building and pier (minimizing work at the summit) and the ground footprint will remain the same. A rendering of the current CFHT facility and a conceptual rendering of the MSE facility are shown in Figure 6.

MSE is conducting Design Phase Studies that will lead to a Construction Proposal Review in 2018 – 2019. In addition to Canada, France and Hawaii, work packages for conceptual designs of all major subsystems are currently distributed among institutes and industries in Australia, China, India and Spain. The Science Team has > 100 members distributed across this collaboration and extends to several other nations. New collaborators to the project are encouraged to participate in both the science and technical aspects of MSE design.

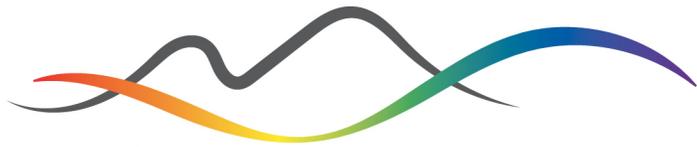

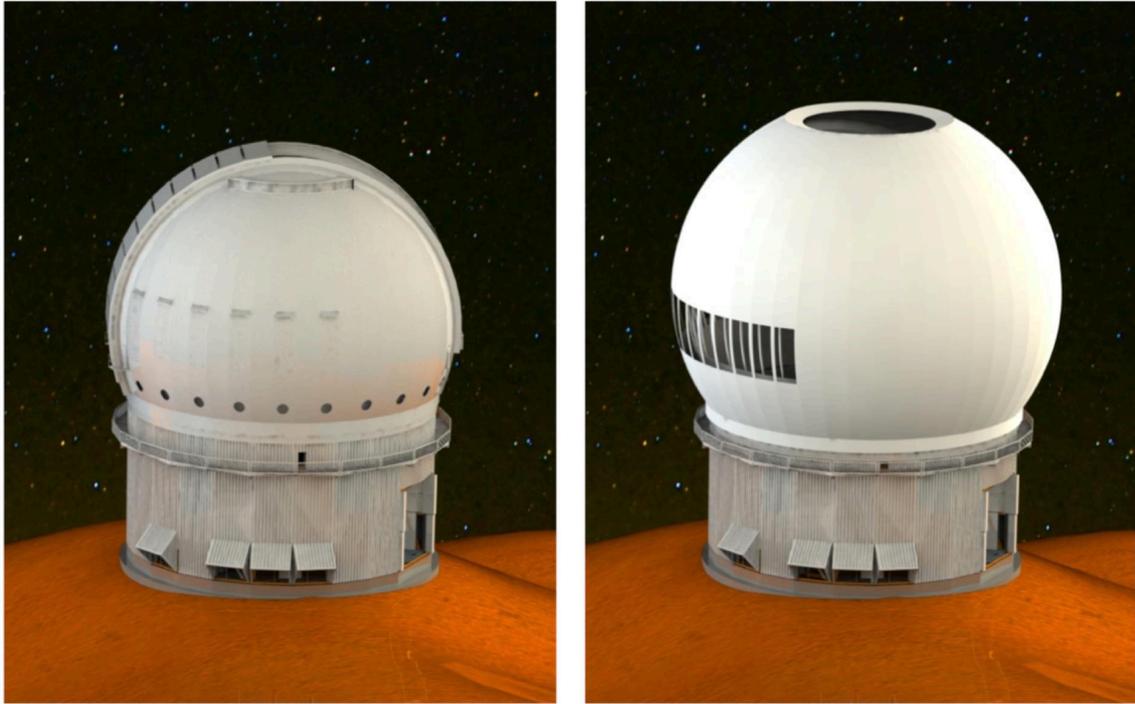

Figure 6: The left panel shows a rendering of the current CFHT facility and the right panel shows a conceptual rendering of the MSE facility.

MSE will benefit from CFHT's 40 years of experience on Maunakea and a support staff deeply rooted in the Hawaii Island community. CFHT/MSE is actively engaging the Hawaii community throughout the planning process and is committed to balancing cultural and environmental interests in the design of the observatory and the organization of the new partnership. With their support, MSE will begin construction once it has passed the Construction Proposal Review and the MSE partnership approves funding, and once all necessary permitting and legal requirements have been fulfilled. This includes the approval of a new master lease for the Mauna Kea Science Reserve.